# Thermodynamics of a chemical reaction model for the atom-field interaction in a three-level laser

Peter Muys

Laser Power Optics

Ter Rivierenlaan 1 P.O. box 4

2100 Antwerpen Belgium

peter.muys@gmail.com

tel +32 477 451916

Abstract- The temporal evolution of the levels in a three-level laser is macroscopically described by a closed set of rate equations. Here, we complement this picture by providing a model which describes through chemical reactions how the levels are evolving.

In most textbooks, the chemical potential is introduced as a concept in quantum statistics. In this paper, we alternatively base its definition on population densities of the excited states in a dopant atom. Then the chemical reaction model delivers a clear and intuitive framework to define the further thermodynamic characteristics of the atom-field interactions such as the chemical potential of the photon and the photon entropy.

**1. Introduction**

In textbooks on quantum statistical physics, the chemical potential is introduced as characteristic parameter during the derivation of the Bose-Einstein distribution for a grand-canonical ensemble of bosons. The boson system is then characterized by two parameters: its absolute temperature and its chemical potential. The temperature is a consequence of the conservation of energy, while the chemical potential is a consequence of the conservation of the number of bosons. The next step is then to argue that photons are bosons and form a photon gas. The gas is thought enclosed in an opaque box or cavity. The chemical potential of the photons then appears as their free (Helmholtz) energy per particle in the photon gas. Most of these textbook approaches stay highly academic, making the concept of chemical potential a hardly explored topic in applied optics and laser physics. As a direct consequence, the physical meaning of the chemical potential of the light remains vague and its potential practical importance for laser engineering is neither clear nor transparent. In [Wurfel] the chemical potential of thermal radiation and of fluorescent radiation from a p-n semiconductor diode is discussed, together with a brief outlook on what the chemical potential of laser radiation could be.

In this paper, we try to work around these conceptual barriers and follow another strategy. We still consider a box, which confines a fluorescent medium. The medium is diluted in a host material, such as methanol in case of a dye, or a crystal in case of a rare-earth element. The fluorescent medium is then called the dopant. The dopant will show different excited levels, which can be filled by an external pump source. We will describe the interaction between the pump, the fluorescent radiation and the dopant, not by concentrating initially on the properties of the fluorescent radiation however. Alternatively, we introduce the chemical potential of light interacting with the fluorescent medium through studying the population densities of the dopant levels and we define the value of the chemical potential as a measure for the strength of the pumping process. Using our concept, the chemical potential of light becomes a sharply defined quantity, having a direct physical interpretation as alternative pump strength parameter. Conservation of particles in the grand-canonical ensemble of quantum statistics is here translated into the condition that each pump photon creates a fluorescent photon, hence the pump efficiency is considered here to reach 100%.

Wherever required, the refractive index of a medium is absorbed into the speed of light and is represented by the symbol c.

## 2. The two-level model

First, we consider a pure two-level model for the dopant in thermal equilibrium. This is the model which Einstein used to derive his A and B coefficients for absorption and emission.
Population inversion of the dopant

$$\frac{N_2}{N_1} > 1$$

does not occur in such a pure two-level system (2 is the upper level index, 1 is the ground level index) where the two levels are filled, as a function of temperature, according to the Boltzmann distribution. This is based on an analysis of the rate equations describing the dynamic evolution of the two levels [Loudon]. Note that the two-level system forms a blackbody where photons are continuously absorbed and re-emitted and so is basically not pumped by an external pump source. It is behaving like an adiabatic thermodynamic system enclosed in an isolated box. No photons escape from the box. No energy is conveyed to the box.

The number of photons $\Phi$ in a single mode of the cavity increases due to both stimulated and spontaneous emission from the upper level and decreases due to absorption from the ground level according to:

$$\frac{d\Phi}{dt} = \sigma c N_2 (\Phi + 1) - \sigma c N_1 \Phi \tag{1}$$

where $\sigma$ is the emission cross section [Loudon]. After regrouping, this becomes:

$$\frac{d\Phi}{dt} = \sigma c \left[ (N_2 - N_1)\Phi + N_2 \right] \tag{2}$$

So an alternative formulation of the evolution of the photon rate is that it increases due to optical gain from the population inversion and due to spontaneous emission from the upper level. In steady-state, the right-hand side of Eq.(1) or (2) is set to zero. Working out this equation, the number of photons becomes linked with the populations as

$$\Phi = \frac{N_2}{N_1 - N_2}$$

or inverting this equation:

$$\frac{N_1}{N_2} = \frac{\Phi + 1}{\Phi} \tag{3}$$

So algebraically, because the right-hand side of Eq.(3) is always larger than 1, N1 is also larger than N2, hence no population inversion is possible. Rearranging this equation, and using the Boltzmann distribution for thermal equilibrium of the levels:

$$\Phi = \frac{1}{\frac{N_1}{N_2}-1} = \frac{1}{\exp\left(\frac{h\nu}{k_B T}\right)-1}$$

Now the spectral energy density of the radiation ( in Js/m³) is given by

$$\rho_{spc} = h\nu\, p(\nu)\Phi(\nu) \qquad (4)$$

where $p(\nu)$ is the mode density for unpolarized radiation propagating over the full solid angle of $4\pi$, and given by

$$p(\nu) = 2 \times 4\pi\, \frac{\nu^2}{c^3}$$

Eq. (4) is Planck's classical formula for the spectral density of a blackbody. We have based its derivation here on the photon balance Eq.(1). The obtained result is completely in agreement with the result normally obtained based on the rate equations for the populations (detailed balance)

$$\frac{dN_2}{dt} = -A_{21}N_2 + B_{12}\rho_{spc}N_1 - B_{21}\rho_{spc}N_2 = -\frac{dN_1}{dt}$$

The two-level model is hence a good starting point to appreciate the duality between a "photon picture" and a " population picture" to study their mutual interaction.

### 3. The three-level model

A two-level system delivers thermal radiation. A three-level system however will emit photoluminescent radiation, which will be characterized by a temperature and an extra parameter: a chemical potential. We will analyse in this section how this thermodynamic concept can be worked out in terms of laser parameters.

In a three-level medium, we add a third level with energy $E_3$ and population density $N_3$ and call it the pump level. It is situated slightly above the upper level $E_2$ with population density $N_2$. We will suppose the pump level is filled by absorbing optical radiation from an external pump source. Population inversion becomes possible now between the upper and the ground level, since we de facto introduce an extra up-transition channel to fill the upper level, by providing for an external pump. There is practically no down-transition from the pump to the ground, because the non-radiative relaxation to the upper laser level is much faster than the radiative de-excitation to the ground level. So there is no stimulated pump radiation present. This pump now fills the pump level. In other words, the pump level population nearly immediately relaxes down to the upper level which is having on its turn a very long lifetime (since it should be a metastable level of the dopant). The upper level then de-excites to the ground level by sending out a photon with energy E

$$E = h\nu_{21}$$

This photon can be either spontaneous or stimulated. Vice versa, the ground level can absorb photons with energy E. Due to the very fast nonradiative relaxation of the pump level to the upper level, we can safely adopt the approximation

$$N_T = N_1 + N_2 + N_3 \approx N_1 + N_2$$

where $N_T$ is the dopant concentration into the host. The energy $h\nu_{32}$ liberated during the relaxation process is dumped as heat into the host material and is called the quantum defect.

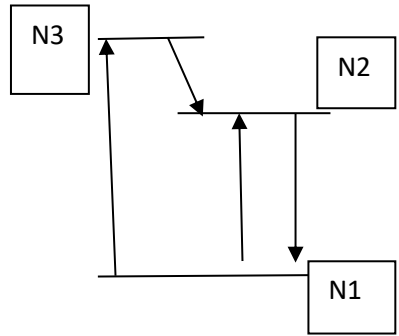

In a three-level system, population inversion becomes possible because it is an open system. This means that energy from an external source (i.e. the pump) is continuously provided to the dopant. All this can be cast into a rate equation model of the level populations and of the photon density, but this is outside the scope of this paper.

In an attempt to implement also Boltzmann statistics to the inverted level 2, one can only mathematically fulfil this condition, by adopting a negative temperature $T_{rad}$ in

$$\frac{N_2}{N_1} = \exp(\frac{-E}{k_B T_{rad}})$$

This is a debatable operation since a negative absolute temperature is unphysical [Geusic]. However, when pumping does result in a population inversion, it can be written in Boltzmann-like form with a *positive* temperature, by introducing an extra parameter M, if we write it as

$$\frac{N_2}{N_1} = M.\exp\left(\frac{-E}{k_B T_{amb}}\right) \qquad (5)$$

with M>0. $T_{amb}$, the ambient temperature, now is also a positive number. The important point here is that we do not need to define an effective temperature of the radiation field responsible for the pumping. We keep the ambient temperature constant all the time and start pumping by gradually increasing the value of M. Its starting and minimum value is 1, resulting in the value of the usual Boltzmann factor i.e. when the system is initially in thermal equilibrium. When pumping starts, the $N_2$ level is gradually filling up, while it is simultaneously emptying due to spontaneous and stimulated fluorescence. We increase the pumping until the medium gets transparent at $N_1=N_2$ and thermodynamically speaking, the light can from now on produce work since a gain is now realised. At this point, we can think of the dopant/host being contained inside an optical resonator which resonates at the frequency $\nu_{12}$. If, under these circumstances, we now further increase the gain, it will ultimately compensate the cavity losses, and the system would be able to start lasing. This is called the threshold condition. Stimulated emission starts to dominate from here and , this point can be proved using the rate equations of the level populations, these population densities clamp at their threshold value if we further increase the pumping above its threshold value. So the factor M also clamps and reaches a maximum value which can be calculated by solving the steady-state rate equations.

We can redefine the factor M now by introducing a quantity $\mu$, having the dimensions of energy, and given by

$$M = \exp(\frac{\mu}{k_B T_{amb}}) \qquad (6)$$

$\mu$ is called the chemical potential. We do not call it yet the chemical potential *of light*, since until now we only considered the population of the levels of the dopant.

Eq.(6) represents a crisp, correct and intuitively clear definition of the concept of chemical potential. We shall show in the subsequent paragraphs that we will be able to deduce all the important thermodynamic characteristics of the radiation interacting with the three-level system, without needing the advanced concepts of the quantum statistics of this radiation.

By substituting Eq.(6) in Eq.(5), we write the chemical potential as:

$$\mu = h\nu_{21} + k_B T_{amb} \ln \frac{N_2}{N_1} \qquad (7)$$

$\mu$ is determined by the ratio of the population densities, which depend on the strength of pumping. Pumping will increase the upper level population, and decrease the lower level population. The minimum value of the chemical potential is zero , corresponding to the unpumped situation: thermal equilibrium given by the Boltzmann factor at ambient temperature. Its maximum value is its value at the laser threshold, due to the clamping of the population densities to their threshold values. Intermediate, its value equals $h\nu_{21}$, at transparency when $N_1 = N_2$ .

**The chemical potential is hence a dynamic variable, depending on the strength of pumping, hence depending on the degree of inversion in the dopant**. We have introduced here the chemical potential , not by Bose-Einstein quantum statistics of the light field, but indirectly, through the excitation of the dopant. Here, we can note the conceptual difference of describing the strength of pumping by the inversion, which is the *difference* of the populations, or now alternatively by the chemical potential, which is determined by the *quotient* of the populations.

### 4. Chemical reaction model of the two-level system

Let us return first in this section to the two-level system. Further insight can be gained by taking advantage of the terminology used in chemical thermodynamics, where the absorption by the dopant A of a photon γ with energy $h\nu_{21}$, resulting in the excited state A* and the process is written as a chemical reaction:

$$A + \gamma \rightleftarrows A^*$$

A corresponds to the lower state 1, with population density $N_1$, A* corresponds to the excited state 2, with density $N_2$. Each state i of the dopant is allocated according to the usual rules of chemical thermodynamics a chemical potential $\mu_i$ depending on the concentration of particles in the state i, according to

$$\mu_i = \mu_{i,0} + k_B T_{amb} \ln \frac{N_i}{N_T} \qquad i=1,2 \qquad (8)$$

where $\mu_{i,0}$ is a conveniently chosen energy reference value. Now, chemical equilibrium of the reaction is established when the sum of the chemical potentials of the reactants equals the sum of the chemical potentials of the reaction products:

$$\mu_1 + \mu = \mu_2$$

where $\mu$ now is by definition the chemical potential of the light interacting with the dopant. If we also require that

$$\mu_{2,0} - \mu_{1,0} = h\nu_{21} \qquad (9)$$

then the expression for the chemical potential of light in a two-level system takes the form

$$\mu = h\nu_{21} + k_B T_{amb} \ln \frac{N_2}{N_1}$$

which is identical to Eq.(7). Note that $N_T$ dropped out of the equation. However, at thermal equilibrium, Boltzmann statistics is followed by the dopant:

$$\frac{N_2}{N_1} = \exp(\frac{-h\nu_{21}}{k_B T})$$

which implies, by substituting this expression into Eq.(7), that the chemical potential of thermal photons is zero. This theorem is here derived in a simple and intuitive way, because we used the population densities and not the photon statistics. In most textbooks, the concept of chemical potential stays elusive, because of the huge prerequisite of the machinery of quantum statistics.

## 5. Chemical reaction model of the three-level system

The fluorescence, the pumping and the relaxation steps are in this case given by the three reactions

$$N_1 + \gamma \rightleftarrows N_2$$
$$N_1 + \gamma_p \rightarrow N_3$$
$$N_3 \rightarrow N_2 + \gamma_h$$

$\gamma_h$ is the quantum of heat deposited in the host material. The overall reaction is found by adding the three partial reactions:

$$2N_1 + \gamma + \gamma_p \rightarrow 2N_2 + \gamma_h \tag{10}$$

We see here that $N_3$ drops out of the equation and is not present in the overall reaction, which means in chemical terminology that it played the role of a catalyst. Moreover, it is interesting to note that the factors 2 in front of $N_1$ and $N_2$ imply that in the level scheme, two arrows should be pointing upwards from $N_1$ and two arrows should be pointing down from $N_2$. While the first statement is directly clear from the graph of the level scheme, the second is not obvious. **The chemical reaction picture dictates that there must be *two* different de-excitation mechanisms for the $N_2$-level. These can be identified of course as the stimulated and spontaneous emission.**

The thermodynamic condition for chemical equilibrium now becomes, based on the overall reaction Eq.(10):

$$\mu + \mu_p - \mu_h = 2(\mu_2 - \mu_1) \tag{11}$$

If we take again as definition for the chemical potential of the fluorescent light $\mu$ the one we found for the two-level system, i.e. Eq.(7), and if we take an analogous expression for the chemical potential of the pump light $\mu_p$

$$\mu = h\nu_{21} + k_B T \ln \frac{N_2}{N_1}$$
$$\mu_p = h\nu_{31} + k_B T \ln \frac{N_3}{N_1} \tag{12}$$

then, substituting these expressions in Eq.(7), we find for the chemical potential of the energy dumped into the host

$$\mu_h = h\nu_{32} + k_B T \ln \frac{N_3}{N_2}$$

where we used Eq.(9).
Taking into account Eq.(4) for the chemical potential of the levels, the following relations are valid:

$$\mu = \mu_2 - \mu_1$$
$$\mu_p = \mu_3 - \mu_1$$

and finally:

$$\mu = \mu_p - \mu_h$$

In thermodynamics, it is proven that the link between the internal energy U, the Helmholtz free energy F and the entropy S' of the cavity is given by

$$F = U - TS' \tag{13}$$

However, we consider not the total cavity values but instead the values of these quantities on a per-particle base, then [Markvart] the thermodynamic definition of the chemical potential as the free energy per particle becomes

$$\mu = \frac{\partial F}{\partial N} \tag{14}$$

and of entropy per particle

$$S = \frac{\partial S'}{\partial N}. \tag{15}$$

It should be noted that these two definitions are *differential* quotients, and not *regular* quotients like F/N or S'/N (as was done in [Wurfel]). So strictly speaking, **S is not the entropy per photon, but rather the change in entropy per photon added to or extracted from the cavity**. Also [Graf] indicates that the ratio is a differential quotient. But nevertheless, S is being called in the literature colloquially the entropy per photon though.

So we find by combining Eqs.(13), (14) and (15) for the chemical potential of the fluorescent light:

$$\mu = h\nu - TS \tag{16}$$

**Conceptually, we now see that the chemical potential *of light* is an inappropriate expression, since its definition Eq.(16) is based on the photon energy, and on the entropy per particle (photon), and we should rather always speak about the chemical potential *of a photon*.** This means that, reconsidering Eq.(7), we now can also allocate an entropy value to a fluorescent photon, to a pump light photon and to the quantum defect heat Q dumped into the host material at temperature T:

$$S_f = k_B \ln \frac{N_1}{N_2}$$

$$S_p = k_B \ln \frac{N_1}{N_3} \tag{17}$$

$$S_h = k_B \ln \frac{N_2}{N_3} = \frac{Q}{T} = \frac{h\nu_{32}}{T}$$

Because

$$\frac{N_1}{N_2} \frac{N_2}{N_3} \frac{N_3}{N_1} = 1$$

it is easy to prove that the three entropies-per-particle are connected by

$$S_p = S_f + S_h \tag{18}$$

The Second Law corrects this relationship to:

$$S_p \leq S_f + S_h \tag{19}$$

since the entropy can only increase during the pumping process. This form of the Second Law was also given in ref[Graf], although based on a different reasoning.

We also see that the thermodynamic condition for transparency of the dopant becomes

$$S_f = 0$$

At transparency, the photon entropy becomes zero and the photon can be qualified as coherent. Above the transparency condition, in case of population inversion, the chemical potential is *larger* than the photon internal energy and the photon entropy becomes negative. The photon now can generate work, meaning that it participates in the amplification process and hence is stimulated. Also note that using our method, we arrive at a positive temperature and a negative entropy. This should be compared with the method of allocating a *negative* temperature to the inversion, which then consequently results in a *positive* entropy.

Let us have a closer look to the inequality in Eq.(19).

When the cavity generates predominantly spontaneous emission, the entropy balance graphically looks like (S is taken as the y-axis of the graph):

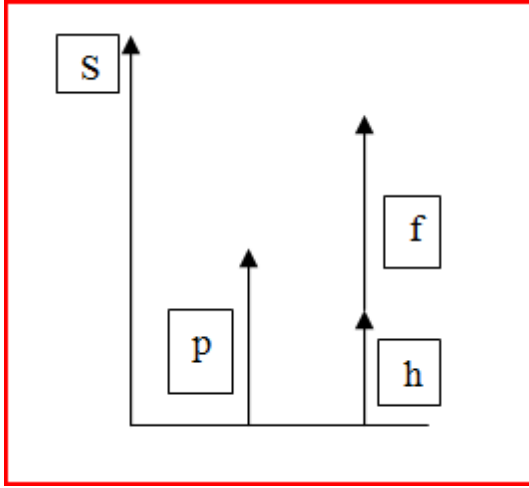

When the cavity generates stimulated radiation, the entropy balance looks like:

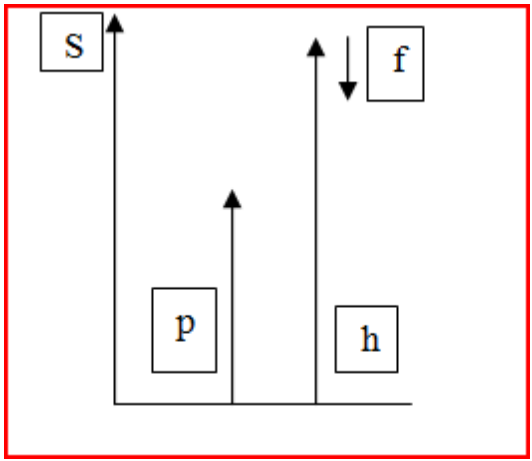

Note the negative value of the stimulated radiation. The Second Law prohibits that the tip of the f-vector would be situated lower on the entropy scale than the tip of the p-vector.

In steady-state, the solutions of the rate equations for the population densities are given by

$$N_2 = \frac{N_T + N_{th}}{2}$$

$$N_1 = \frac{N_T - N_{th}}{2}$$

where $N_{th}$ is the inversion at threshold:

$$N_{th} = (N_2 - N_1)_{th} = \frac{1}{\sigma c \tau_c}$$

We see that these expressions are independent of the strength of the pump, and moreover also independent of the photon density.

$$S_f = -k_B \ln \frac{N_2}{N_1} = -k_B \ln \frac{\frac{N_T}{N_{th}} + 1}{\frac{N_T}{N_{th}} - 1}$$

This logarithm can be expanded in the quickly converging power series

$$S_f = -2k_B \left[ \frac{N_{th}}{N_T} + \frac{1}{3}\left(\frac{N_{th}}{N_T}\right)^3 + \frac{1}{5}\left(\frac{N_{th}}{N_T}\right)^5 + \ldots \right]$$

The pump entropy can be calculated, based on the rate equation

$$\frac{dN_3}{dt} = \sigma_p c N_D N_1 - \frac{N_3}{\tau_{pu}}$$

where $N_D$ is the photon density of the diode laser pump and $\tau_{pu}$ is the relaxation time of the pump level to the upper laser level. We have assumed here that there is no stimulated radiation back from the pump level to the ground level. In steady-state, this gives

$$\frac{N_3}{N_1} = \tau_{pu} \sigma_p c N_D \triangleq R$$

where R is defined as the pump strength. So the entropy per pump photon is given by taking the logarithm of this expression:

$$S_p = -k_B \ln R$$

Now, using Eq.(), we hence find for the entropy from the quantum defect in the host

$$S_h = S_p - S_f = k_B \left( \frac{2N_{th}}{N_T} - \ln R \right)$$

Let us now check if our definition of radiation entropy based on population densities, corresponds with the radiation entropy based on photon statistics in the case of a blackbody. The average number of photons per thermal mode is given by the Planck formula

$$\bar{n} = \frac{1}{\exp\left(\frac{h\nu}{k_B T}\right) - 1}$$

Reworking this, we find:

$$\frac{\bar{n} + 1}{\bar{n}} = \exp\left(\frac{h\nu}{k_B T}\right) \qquad (20)$$

In thermal equilibrium, the population densities follow the Boltzmann distribution

$$\frac{N_1}{N_2} = \exp\left(\frac{h\nu}{k_B T}\right) \quad (21)$$

By substituting this expression in eq.(9), we find for the entropy of a thermal photon:

$$S_f = \frac{h\nu}{T} \quad (22)$$

Because the right-hand sides of Eqs.(20) and (21) are equal, we basically find back Eq.(3), which links the populations to the number of photons.

Eq.(22) can be compared with the expression of [Graf], who first calculated the entropy of the thermal mode,

$$S' = k_B\left[(\bar{n}+1)\ln(\bar{n}+1) - \bar{n}\ln(\bar{n})\right]$$

and then used Eq.(15):

$$\sigma_{th} = \frac{dS'}{d\bar{n}} = k_B \ln\frac{\bar{n}+1}{\bar{n}} = \frac{h\nu}{T}$$

By the way, this also clearly shows that

$$\frac{dS'}{d\bar{n}} \neq \frac{S'}{\bar{n}}.$$

Extending Eqs.(20,21) to fluorescent photons having a chemical potential, i.e. using Eq.(5), the entropy per fluorescent photon now is given by

$$S_f = \frac{h\nu - \mu}{T}$$

which also follows directly from Eq.(16).

### 6. The photon distribution function for a three-level laser

Eq.(1) describes the photon balance, in case there is no output coupling from the cavity. When output coupling can take place, through a partial reflector as one of the cavity mirrors, the photon balance for a single mode in the cavity in steady-state becomes

$$\sigma c N_2(\Phi + 1) - \sigma c N_1 \Phi = \frac{\Phi}{\tau_c}$$

This determines the number of photons in the mode as

$$\Phi = \frac{N_2}{\frac{1}{\sigma c \tau_c} - N_2 + N_1} = \frac{N_2}{N_{th} - N_2 + N_1} = \frac{1}{\frac{N_{th}}{N_2} - 1 + \frac{N_1}{N_2}} \quad ()$$

We now here introduce the chemical potential for the ratio of the population densities, as given by Eq.() and become:

$$\Phi = \frac{1}{\exp\left(\frac{h\nu - \mu}{k_B T}\right) - 1 + \frac{N_{th}}{N_2}}$$
()

So eq() makes visible how the Bose-Einstein distribution for a closed cavity

$$\Phi = \frac{1}{\exp\left(\frac{h\nu - \mu}{k_B T}\right) - 1}$$

is changed by the output coupling from the cavity. By defining a new parameter $\varepsilon$

$$\varepsilon \triangleq \frac{N_{th}}{N_2} = \frac{2N_{th}}{N_T + N_{th}}$$

the expression for the chemical potential becomes

$$\mu = h\nu - k_B T \ln\left(\frac{1}{\Phi} + 1 - \varepsilon\right)$$

Or, using the relation Eq.() between chemical potential and entropy per photon:

$$S_f = k_B \ln\left(\frac{1}{\Phi} + 1 - \varepsilon\right)$$

Finally, Eq.() can be written alternatively as

$$\frac{N_1}{N_2} = \frac{\Phi + 1}{\Phi} - \varepsilon$$

## 9. Summary

In this paper, we have based the definition of chemical potential not on photon statistics, but on population densities. Briefly stated, the chemical potential is an alternative way of specifying the strength of pumping. In a next step, we have put together a chemical reaction model of a three-level dopant. Following this strategy, we were able to develop a thermodynamic model of the laser, including
-the chemical potential of a photon
-a simple and elegant proof of the fact that the chemical potential of a thermal photon is zero
-the chemical potential and the entropy of a fluorescent photon
-the entropy of a thermal photon
-the entropy of a laser photon
A photon is characterized by a number of parameters. These are classically its energy, its frequency, its momentum and its polarization. To complete the picture, we now can add its entropy or equivalently its chemical potential. Finally, this defines also the coherence of the photon: a photon is coherent as soon as its chemical potential is higher than its internal energy.